\documentclass[preprint2]{aastex} 
 
\shortauthors{Forbrich et al.} 
 
\begin{document} 
 
 
\title{High-Precision Radio and Infrared Astrometry of LSPM J1314+1320AB
-- I: Parallax, Proper Motions, and Limits on Planets}                  
 
 
\author{Jan Forbrich\altaffilmark{1,2}, Trent J. Dupuy\altaffilmark{3},
Mark J. Reid\altaffilmark{2}, Edo Berger\altaffilmark{2}, Aaron Rizzuto\altaffilmark{3},
Andrew W. Mann\altaffilmark{3}, Michael C. Liu\altaffilmark{4}, Kimberly
Aller\altaffilmark{4}, Adam L. Kraus\altaffilmark{3}} 
 
 
\altaffiltext{1}{University of Vienna, Department of Astrophysics,
T\"urkenschanzstr. 17, 1180 Vienna, Austria}                            
\altaffiltext{2}{Harvard-Smithsonian Center for Astrophysics, 60
Garden St, Cambridge, MA 02138, USA}                                    
\altaffiltext{3}{The University of Texas at Austin, Department of
Astronomy, 2515 Speedway C1400, Austin, TX 78712, USA}                  
\altaffiltext{4}{Institute for Astronomy, University of Hawai'i,
2680 Woodlawn Drive, Honolulu, HI 96822, USA}                           

 
\begin{abstract} 
We present multi-epoch astrometric radio observations with the Very
Long Baseline Array (VLBA) of the young ultracool-dwarf binary LSPM
J1314+1320AB .   The radio emission comes from the secondary star.
Combining the VLBA data with Keck near-infrared adaptive-optics observations
of both components, a full astrometric fit of parallax 
($\pi_{\rm abs}=57.975\pm0.045$~mas, corresponding to a distance
of $d=17.249\pm0.013$~pc), proper motion ($\mu_{\rm \alpha cos \delta}=-247.99\pm0.10$~mas\,yr$^{-1}$,
$\mu_{\delta}=-183.58\pm0.22$~mas\,yr$^{-1}$), and orbital motion
is obtained. Despite the fact that the two components have nearly
identical masses to within $\pm2$\%, the secondary's radio emission
exceeds that of the primary by a factor of $\gtrsim$30, suggesting
a difference in stellar rotation history, which could result in different magnetic field configurations.
Alternatively, the emission could be anisotropic and beamed toward us for the secondary
but not for the primary.  Using only reflex motion, we exclude planets
of mass 0.7 to 10~$M_{\rm jup}$ with orbital periods of 600 to 10
days, respectively.   Additionally, we use the full orbital solution
of the binary to derive an upper limit for the semi-major axis of
0.23~AU for stable planetary orbits within this system.  These limits
cover a parameter space that is inaccessible with, and complementary
to, near-infrared radial velocity surveys of ultracool dwarfs.  
Our absolute astrometry will constitute an important test for the
astrometric calibration of {\it Gaia}.                                        
 
\end{abstract}

 

\section{Introduction} 
 
Since the discovery of nonthermal radio emission from ultracool dwarfs \citep{ber01}, where following \citet{kir97}, the term ultracool dwarfs here refers to spectral
types M7 or later, and the 
detection of a nearby ultracool dwarf using Very Long Baseline Interferometry
(VLBI; \citealp{fbe09}), the prospect of using absolute VLBI astrometry
to measure parallax and proper motion, and to search for
residual reflex motions due to otherwise unseen planets, has been
intriguing.                                                             
Additionally, precise astrometry that can reveal the orbital evolution of nearby
young multiple systems is of particular interest, since complete orbit
solutions are a reliable method to determine masses, which can be used to test
theoretical models of substellar systems \citep{bar15,dup15}. Only a few young binaries
with very low-mass components currently have reliable mass estimates.
 
While planets around M~dwarfs are thought to be ubiquitous (e.g.,
\citealp{dre15}), the search for them is increasingly difficult for ultracool dwarfs. 
Planet searches toward stars of very low mass are usually carried out using
transit or radial velocity (RV) techniques. For ultracool dwarfs,
RV precisions of only 100--300~m\,s$^{-1}$ can be reached when using high-resolution
near-infrared spectra of slowly rotating stars \citep{bla10}, and the precision 
is lower for rapidly rotating stars.
Given the difficulties of RV monitoring, astrometric observations
provide an attractive alternative.                                      
 
With our first VLBI detection of an ultracool dwarf \citep{fbe09},
we pointed out the feasibility of looking for orbiting extrasolar planets
by detecting their reflex motion with VLBI
astrometry.  The first astrometric monitoring campaign toward TVLM 513-46546
confirmed the feasibility with significant upper limits on the presence of
extrasolar planets \citep{fbr13}, excluding (at a significance
of 3$\sigma$) planets of masses and orbital periods ranging from
3.8~$M_{\rm jup}$ with an orbital period of 16 days to 0.3~$M_{\rm
jup}$ with an orbital period of 710 days.  Previously, such searches
had been carried out for early M-type dwarfs (e.g., \citealp{bow09}).
The principal advantage of VLBI observations would to
obtain absolute parallax and proper motions (tied to an extragalactic
reference frame).  Astrometric
observations can also be carried out in the optical wavelength range
with 10m-class telescopes, but require conversion of relative to absolute parallax
(e.g., \citealp{sah16}).                    
 
\begin{figure*} 
\epsscale{2.2}
\plotone{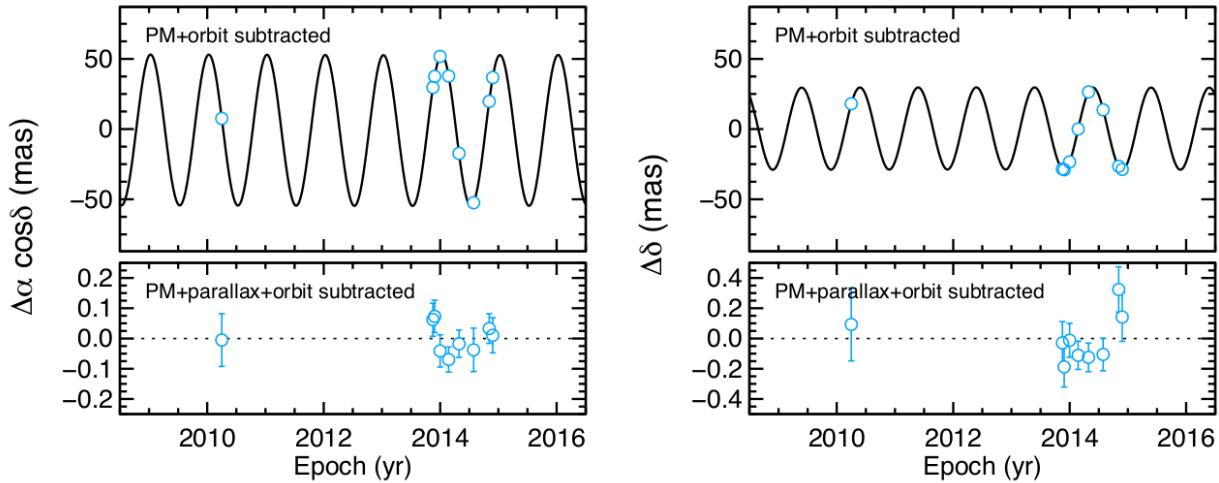}
\caption{Astrometric residuals in RA ($\alpha$cos$\delta$, left panel) and Declination ($\delta$, right panel). The upper plots show the effect of
the parallax, when proper motion and orbital motion are subtracted. The lower
plots show the astrometric residuals when additionally subtracting the
parallax. The error bars indicate 1$\sigma$ errors.\label{fig_1dres}}   
\end{figure*} 
 
\citet{sch14} presented LSPM J1314+ 1320AB as a potential ``benchmark'' binary for the determination of substellar masses to calibrate evolutionary models at young ages. Designations of primary and secondary components are based on
infrared magnitude. This system has a high proper motion (e.g., \citealp{lep05}). 
\citet{law06} estimated
the spectral type as a late M~dwarf from multi-band photometry and
detected a companion at a separation of $0\farcs$13 from optical
imaging.  An optical astrometric parallax measurement was obtained
by \citet{lep09} who found $\pi_{\rm abs}=61.0\pm2.8$~mas
($d=16.39^{+0.79}_{-0.72}$~pc) and a relative proper motion of $\mu_\alpha=-243$~mas\,yr$^{-1}$
and $\mu_\delta=-186$~mas\,yr$^{-1}$ (no errors given). \citet{lep09}
also obtained an optical spectrum to determine the unresolved spectral
type as M7.0e with strong H$\alpha$ emission.                           
 
At about the same time, we discovered a bright radio counterpart
to LSPM J1314+1320AB with the Karl G. Jansky Very Large Array (VLA)
of the National Radio Astronomy Observatory (NRAO), with the first
observations reported in \citet{mcl12}. Given its brightness, LSPM
J1314+1320AB appeared to be a good candidate for VLBI radio observations.
Indeed, \citet{mcl11} reported a VLBI detection of LSPM J1314+1320AB
in the 3.5~cm band, noting that just one of the two components of
this binary appeared to show radio emission. With just one epoch
of VLBI data, it was impossible to tell which of the two components
had been detected. The same paper also reported the results of multi-frequency
VLA observations and optical monitoring, confirming a persistent
radio counterpart ($\gtrsim$1~mJy in the 3.5~cm
band) and revealing variability with a period of about four hours.
Most recently, \citet{wil15} reported {\it Chandra} and {\it Swift}
X-ray detections in a multi-wavelength observation campaign. The
angular resolution of the associated VLA observations was not sufficient
to resolve the binary, but new radio variability patterns emerged
from these observations, with slow variability overlaid by short
flares lasting only a few minutes.                                      
 In a near-infrared spectroscopic study of LSPM J1314+1320AB, \citet{sch14}
concluded that the system has an age of $\sim30-200$~Myr, and their
preliminary mass estimates indicated that both components are near
the substellar boundary.                                                
 
Given the consistently detectable radio emission
and one successful VLBI detection, we selected LSPM J1314+1320AB
as a target for VLBI astrometric monitoring with the goal of measuring
parallax and proper motion and
then look for reflex motions caused by extrasolar planets. 
In a highly synergistic combination of datasets, we have analyzed
our VLBI astrometric data together with Keck near-infrared adaptive-optics
monitoring, combining the advantages of absolute radio astrometry
(even if on just one component) with relative infrared astrometry
of both components (\citealp{dfr16}, hereafter Paper~II).                         
The layout of this paper is as follows. In Section~\ref{sec_obs},
we present the VLBI observations and data analysis. The contemporaneous
Keck observations, the full astrometric fit, and all its parameters
are presented in Paper~II. In Section~\ref{sec_rad}, we discuss the
results of the VLBI astrometry, the radio properties of the two binary
components, and prospects for comparison with {\it Gaia} data. In Section~\ref{sec_pla},
we search for residual reflex motions of extrasolar planets,
after taking into account parallax, proper motion, and binary orbital
motion. Finally, in Section~\ref{sec_sum}, we summarize our conclusions.

\section{Observations\label{sec_obs}} 
 
\begin{figure} 
\epsscale{1.0}
\plotone{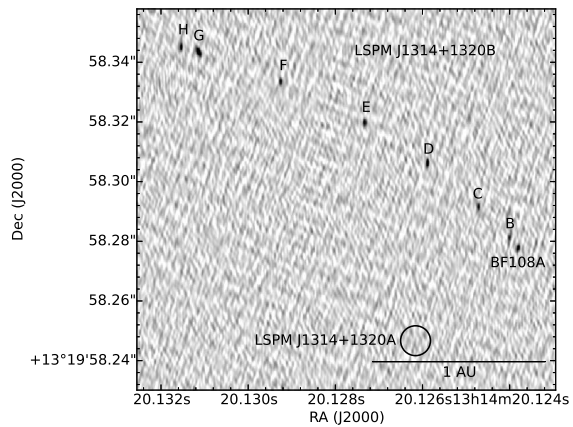}
\caption{Coadded images of VLBA epochs BF108A-H, centered on the
predicted position of the primary component LSPM J1314+1320A. The
primary remains undetected; its position is marked with a circle
of radius 5~mas. All eight detections of the secondary, starting
with the westernmost point, are clearly visible, and the shape of
the orbit is recognizable (see Fig.~1 in Paper~II).\label{fig_merged}}  
\end{figure} 
 
Observations were obtained with NRAO's Very Long Baseline Array (VLBA) under project
BF108.  The target was observed over a total of eight epochs, each with a duration 
of six hours. The observing log
is listed in Table~\ref{tab_obslog}. At the beginning and end of
each epoch, ``geodetic block'' observations toward a selection of sources in
the International Celestial Reference Frame catalog were carried out to improve
the astrometric accuracy \citep{reb04}. 
These were done in dual-band mode with eight S-band channels
and eight X-band channels.

In the main science block,
rapid-switching phase-referencing observations were carried out using
the nearby phase calibrator J1309+1154 at an angular separation of
$1\fdg8$ from the target, with 40~seconds on target and
40~seconds on J1309+1154.  The total on-source time
for the target was 2.3~h per epoch.                                     
The observations were recorded with a bit rate of 2.048~Gbps, using
the Roach Digital Backend and the polyphase filterbank digital signal-processing
algorithm. The X-band setup for the science observations consisted
of 16 channels with a bandwidth of 32~MHz each, covering the frequency
range of 7.052--7.324~GHz in dual polarisation.  The initial correlator position was estimated from earlier observations and then refined for later epochs.            
 
When planning the sequence of observations, it was challenging to obtain
an optimal temporal spacing of the eight observing epochs.  Contrary
to astrometric observations designed to obtain an optimal parallax
measurement, our observations also had to consider the detection 
of astrometric reflex motions on an unknown timescale.  In order to cover 
a wide range of epoch spacings (to probe a range of possible orbital
periods), we scheduled the observations in multiples
of eleven days near day 0, 11, 44, 99, 165, 242, 352, and 374, starting
on 2013 November 16.  However, external scheduling constraints 
meant that this sequence was not exactly realized (see Table~\ref{tab_obslog}).

\begin{deluxetable}{llrl} 
\tablecaption{VLBA observing log\label{tab_obslog}} 
\tablewidth{0pt} 
\tablehead{ 
\colhead{Epoch} & \colhead{Date (UTC)} & \colhead{Day} & \colhead{Antennas\tablenotemark{a}}}
\startdata 
BM327  & 02 Apr 2010 02:00--10:00 & --1324 & VLBA10+GBT\\ 
BF108A & 16 Nov 2013 13:30--19:30 &      0 & VLBA9(-FD)\\ 
BF108B & 27 Nov 2013 12:46--18:46 &     11 & VLBA9(-FD)\\ 
BF108C & 31 Dec 2013 10:00--16:00 &     45 & VLBA10\\ 
BF108D & 23 Feb 2014 07:00--13:00 &     99 & VLBA10\\ 
BF108E & 29 Apr 2014 02:40--08:36 &    164 & VLBA9(-NL)\\ 
BF108F & 29 Jul 2014 20:43--02:44 &    255 & VLBA10\\ 
BF108G & 04 Nov 2014 14:15--20:16 &    353 & VLBA9(-MK)\\ 
BF108H & 26 Nov 2014 12:51--18:52 &    375 & VLBA9(-BR)\\ 
\enddata 
\tablenotetext{a}{VLBA10 is the full array, VLBA9 is missing one
antenna as listed, the BM327 observation also involved the Green
Bank Telescope (GBT)}                                                   
\end{deluxetable} 
 
The raw interferometric visibility data were calibrated and analyzed
with the NRAO Astronomical Image Processing System (AIPS)
software package using standard procedures. Initially, ionospheric
delays were removed using measurements of the total electron content
provided by Global Positioning System measurements, and all data were corrected
using updated USNO Earth orientation parameters. Residual delays from
clock drifts and zenith atmospheric delays were measured with the
geodetic blocks and removed from the phase-reference data.
Using observations of the strong calibration source J1800+782, we
removed electronic delays and differences among the IF bands.  
Given the availability of a bright
phase calibrator, we included a final determination of multi-band
delays on this source and applied these to the entire dataset. Then,
we interpolated the phases for all rapid-switching reference sources.
 
After calibration, the visibility data were imaged with the AIPS
task IMAGR, with the Briggs weighting parameter ``robust'' set to zero.
The resulting average synthesized beam given the ($u,v$) coverage
of our observations was non-circular with a size of 2.9$\times$1.2~mas,
with the larger extension in Declination.  At the target distance,
this corresponds to a linear resolution of about 10$\times$4~$R_\odot$.
Since M dwarfs have a radius of $\approx0.1$~$R_\odot$, it is unlikely
that these observations resolve any coronal structures in the target(s). 
However, the centroid of the radio position could still be affected somewhat in the case
of unevenly distributed active regions, as discussed below.             
 
The asymmetric beam means that the absolute positions are better
constrained in RA than in Declination. However, this is not
only governed by the synthesized beam shape. Residual tropospheric
and ionospheric delays generally affect Declination more than RA, and 
astrometric noise is expected to be higher in Declination
than what would be expected from the beam shape alone \citep{reh14}. 
Combined with
the fact that the parallax signature is larger in RA than in Declination,
the RA data dominates the parallax estimate.         
 
To further improve our astrometric fit by increasing the time baseline,
we also show our VLBA observations first reported by \citet{mcl11}.
These were carried out in April 2010 (project ID BM327) with both
the VLBA and the Green Bank Telescope to increase sensitivity. Since
the associated geodetic-block observations were not used when we
previously reported the VLBA detection, we re-reduced the BM327 data
as described above.

\section{The Radio Properties and Motion of LSPM J1314+1320AB\label{sec_rad}}

\subsection{Detections and Flux Densities} 
 
A single unresolved radio source was detected toward the target at
all epochs, with peak flux densities ranging from $477$ to $849$~$\mu$Jy,
with an average of 666~$\mu$Jy.  The detailed fit results, including
absolute positions, nominal position errors, and synthesized beam
sizes are listed in Table~\ref{tab_pos}. These position
errors merely reflect the precision of the fit; they do not reflect
systematic errors, which we will discuss below. While it was not immediately
obvious that the detections always corresponded to the same
component of the binary system, it became clear from the global astrometric
fit (Paper~II) that radio emission is indeed only detected from LSPM
J1314+1320B.                                                            
 
The radio light curves suggest variability on timescales
of weeks to months, but we refrain from discussing variability in
more detail here. Given the known short-timescale variability in
this source and the fact that we need to perform rapid-switching between the
target and calibrator, our coverage of the short
timescales is not ideal (cf. \citealp{wil15}).   
Interestingly, the VLBA flux densities are consistently lower by $\gtrsim20$\%
than the found in the VLA observations. Since we do
not expect to resolve the stellar corona, there are two plausible explanations
for this difference.  The most
likely explanation lies in uncertainties in the VLA/VLBA absolute flux scales,
although noise in the phase-referencing data might scatter flux away from
the target.  Another effect that might contribute to this discrepancy
is that the rapid phase-referencing VLBI observations do not provide
good coverage on short timescales of minutes even though flaring
variability (by factors of $\sim2$) has been observed toward this
source \citep{wil15}. A dedicated study for radio variability at
high time resolution on TVLM 513-46546 revealed even stronger, short
flares \citep{wol14}. Given a low number of flares expected per epoch,
we thus could easily have missed some short flares while observing the
phase calibrator, and they may be bright enough to increase the time-averaged
flux density when observed with the VLA. A detailed assessment would
require simultaneous observations with the VLA and VLBA.                
 
\subsection{Parallax and Proper Motion} 
 
In contrast to our earlier work on TVLM 513-46546 \citep{fbr13},
the astrometric fit for LSPM J1314+1320AB has to take into account
not only the parallax and proper motion, but also the orbital motion
in the binary system.  With just one of the two components detected
in the radio, the combination radio data with with resolved
near-infrared monitoring (Paper~II) provide mutual constraints on the 
positions for both components.          
The VLBA data, with 
much higher (and absolute) astrometric accuracy, determine the
parallax and proper motion, while the
Keck data, with the detection of both stars,
determine the binary orbit and other properties of the system.
In particular, the Keck data constrain the shape of the orbit while
the VLBA data constrain the semimajor axis of the orbit for the secondary,
(a parameter unconstrained by the Keck data alone).                       
The full astrometric fit, including the determination of the
orbit, is discussed in Paper~II. 

In Figure~\ref{fig_1dres},
we show the RA and Declination residuals as a function of time,
subtracting the orbital and proper motion (to highlight the
parallax) and additionally subtracting the parallax (to show residuals
that constrain planetary perturbations).                                
The parallax of LSPM J1314+1320AB is $\pi_{\rm
abs}=57.975\pm0.045$~mas, i.e., with an error of $\leq0.08$\%, corresponding to a distance of $d=17.249\pm0.013$~pc. The
error of 45~$\mu$as is 60 times better than for the
previous parallax determination ($\pi=61.0\pm2.8$~mas,
\citealp{lep09}, which was estimated from a relative measurement).
 Similarly, our constraints on the absolute proper motion improve
on the earlier determination.  We find $\mu_{\rm \alpha cos \delta}=-247.99\pm0.10$~mas
and $\mu_{\delta}=-183.58\pm0.22$~mas. Given the difficulties of
obtaining an absolute proper motion measurement from optical (relative)
measurements, the match to the previous results is very good.

\begin{deluxetable}{rrrrr} 
\tablecaption{2M1314 detections\label{tab_pos}} 
\tablewidth{0pt} 
\tablehead{ 
\colhead{Epoch} & \colhead{RA (J2000)} & \colhead{Dec (J2000)} &
\colhead{Beam size\tablenotemark{a}} & \colhead{Peak flux density}\\    
                & \colhead{13:14:XX (J2000)} & \colhead{+13:19:XX
(J2000)} & \colhead{(mas)}            & \colhead{($\mu$Jy\, bm$^{-1}$)}}
\startdata 
BM327  & 20.1914692 $\pm$ 0.0000013 & 58.970242 $\pm$ 0.000040 &
2.9$\times$1.1 & 525$\pm$17\\                                           
BF108A & 20.1238242 $\pm$ 0.0000009 & 58.277745 $\pm$ 0.000027 &
2.5$\times$1.0 & 748$\pm$24\\                                           
BF108B & 20.1239566 $\pm$ 0.0000014 & 58.273532 $\pm$ 0.000051 &
2.2$\times$1.2 & 477$\pm$25\\                                           
BF108C & 20.1237042 $\pm$ 0.0000010 & 58.267385 $\pm$ 0.000036 &
3.3$\times$1.3 & 592$\pm$23\\                                           
BF108D & 20.1208275 $\pm$ 0.0000008 & 58.270925 $\pm$ 0.000026 &
2.8$\times$1.1 & 747$\pm$21\\                                           
BF108E & 20.1147535 $\pm$ 0.0000009 & 58.271672 $\pm$ 0.000026 &
2.7$\times$1.1 & 728$\pm$22\\                                           
BF108F & 20.1090463 $\pm$ 0.0000012 & 58.219960 $\pm$ 0.000034 &
2.7$\times$1.1 & 531$\pm$21\\                                           
BF108G & 20.1103882 $\pm$ 0.0000010 & 58.135605 $\pm$ 0.000027 &
3.1$\times$1.5 & 849$\pm$22\\                                           
BF108H & 20.1107254 $\pm$ 0.0000009 & 58.123189 $\pm$ 0.000036 &
3.5$\times$1.2 & 659$\pm$23\\                                           
\enddata 
\tablenotetext{a}{FWHM synthesized beam size in mas. The position angle
was always within 5$^\circ$ of the North-South direction, with the
exception of epoch BF108G at 12$^\circ$.}                               
\end{deluxetable}

\begin{figure} 
\epsscale{1.0}
\plotone{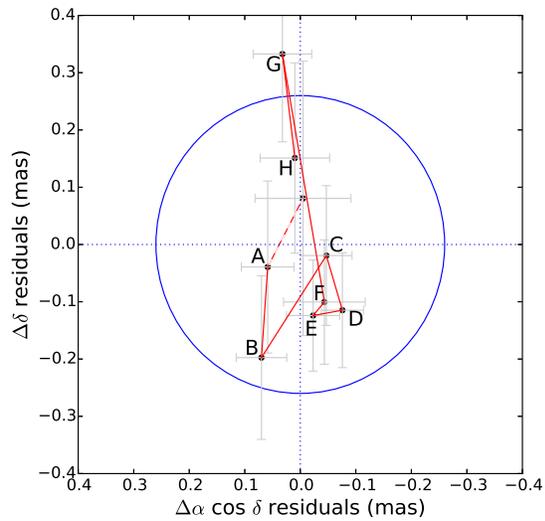}
\caption{2D astrometric residuals of LSPM J1314+1320AB, after parallax,
proper motion, and orbital motion have been subtracted. The error
bars indicate the nominal errors with the error floors added in quadrature.
The blue circle indicates the 3$\sigma$ radial (2D) rms, used as
an upper limit for the presence of reflex motions.\label{fig_2dres}}    
\end{figure} 
 
\begin{figure} 
\epsscale{1.0}
\plotone{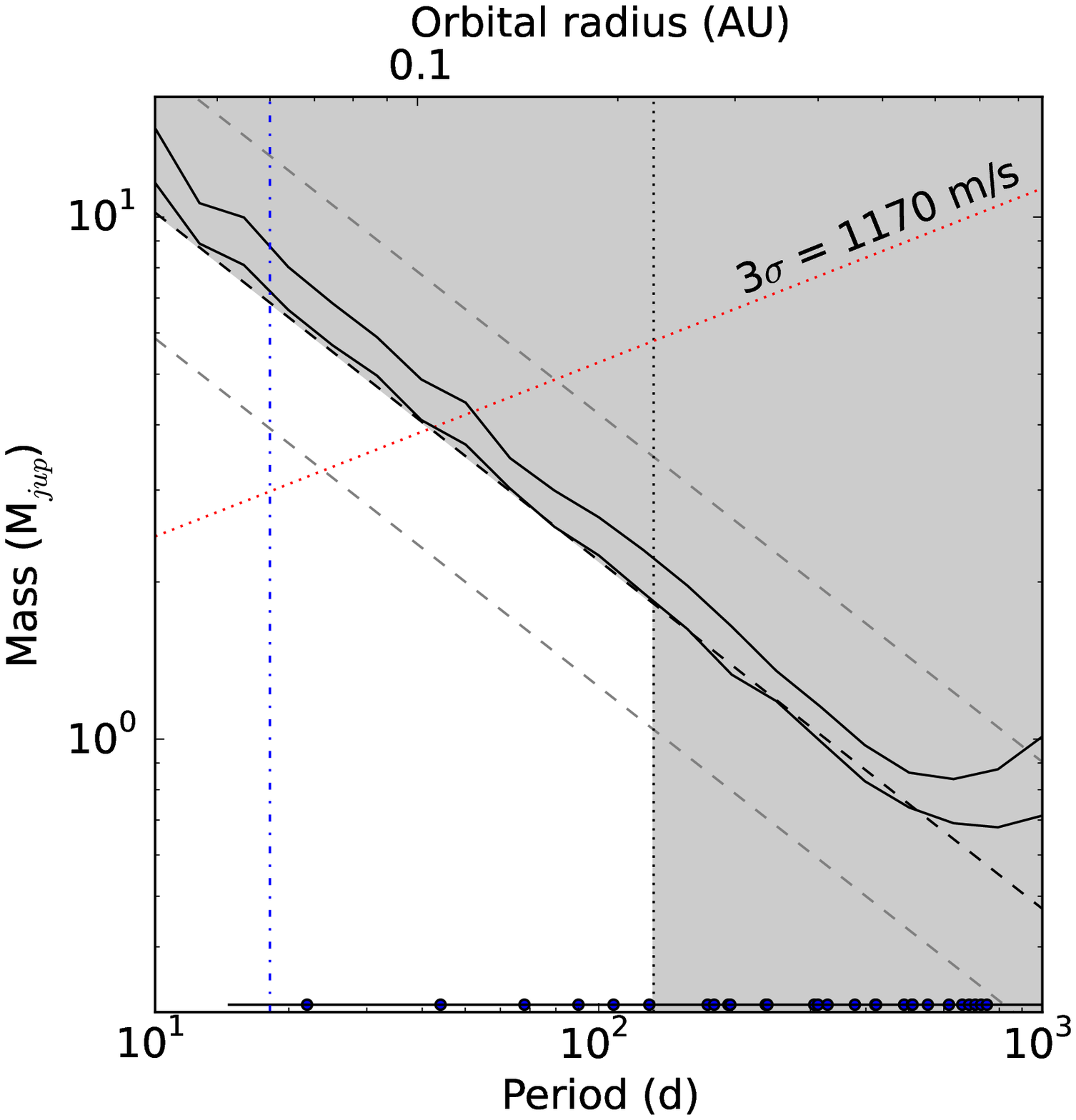}
\caption{Limits to the presence of a hypothetical single extrasolar
planet that would cause reflex motions of LSPM J1314+1320B above
the observed astrometric residuals (see text). The grey areas denote
the parameter space that can be ruled out. The black dashed line
indicates the limit corresponding to the 3$\sigma$ 2D rms of the
residuals, while the gray dashed lines indicate the 3$\sigma$ rms
limits separately in Declination (upper line) and RA (lower line).
Depending on the orbit geometry and in particular the eccentricity,
the limit is between these two extremes. The two black continuous
lines show 95\% detection completeness of simulated orbits that are
circular (lower curve) or have $e=0.6$ (upper curve). As an additional
constraint, we point out that orbits with $P>133$~days are unstable
in this system (based on the orbit solution from Paper~II, see text)
and thus ruled out, too. The area above the red dotted line is the
approximate phase space, with $M$sin$i$ instead of $M$, that could
additionally be ruled out in radial velocity monitoring when assuming
an uncertainty for rapidly rotating stars (3$\sigma$=1170~m\,s$^{-1}$)
based on the work by \citet{bla10}. The blue dots at the bottom of
the plot are shown as a measure of our coverage of parameter space.
Dots are placed at orbital periods $P$ where there is a time baseline
of $P/2$ among the VLBA observing epochs, with lines indicating intervals
[1/4$P$,3/4$P$], i.e., where there is good sensitivity for the maximum
reflex motion caused by such orbits. Finally, the vertical blue dash-dotted
line indicates the outer radius of the habitable zone around the
star according to \citet{sel07}.\label{fig_planet}}                     
\end{figure} 
 
\subsection{Prospects for Cross-checking with {\it Gaia}} 
 
LSPM J1314+1320AB will be an important test case for the {\it Gaia}
mission, as it will be possible to directly compare their astrometric
calibration against the absolute astrometry presented here. The uncertainty
of our absolute parallax measurement is 45~$\mu$as. For a single
star with half of the optical brightness of the combined system LSPM
J1314+1320AB ($V-I_C=3.89$~mag and $V=15.83$~mag, \citealp{lep09}), {\it Gaia} is expected to deliver a nominal end-of-mission parallax uncertainty that is
better by a factor of about two \citep{gaia_sp}, at $\sigma_\pi$=20~$\mu$as. The
main complicating factor that may increase this uncertainty will
be that LSPM J1314+1320AB is, of course, a binary with a separation
of $<0\farcs2$, with about half of the 9.6-year orbit monitored during
the {\it Gaia} mission. The expectation is, however, that {\it Gaia}
will ''systematically resolve double stars down to separations of
$\sim0\farcs1$.''                                    

In a similar test for the previous astrometric mission {\it Hipparcos},
VLBI results confirmed a 10\% error in the combined parallax of the Pleiades 
cluster stars \citep{mel14}.  Distance measurements
derived with several other techniques had also disagreed with the {\it Hipparcos}
result.   The discrepancy had been large enough to imply severe problems for
astrophysical models of young star stellar evolution.  Thus, direct VLBI 
measurement of the absolute parallax can be of great value in the {\it Gaia} era.

For brown dwarfs, which are fainter and cooler than LSPM J1314+1320AB, the
astrometric performance of the VLBA will be superior to that of {\it Gaia},
whenever the target shows sufficiently bright radio emission. In the absence of
a correlation between the optical brightness of a brown dwarf and its radio
emission, this is best shown in a concrete example. For LSPM J0036+1821, an L3.5
dwarf with radio emission \citep{ber02}, the {\it Gaia} astrometric performance
model predicts an uncertainty of the parallax measurement of $\sigma_\pi$=115~$\mu$as, more than twice the uncertainty that we have reached
for LSPM J1314+1320AB.

\subsection{Upper Limits to the Radio Emission of LSPM J1314+1320A} 
 
While the first VLBA observations of \citet{mcl11} produced just
one radio counterpart toward LSPM J1314+1320AB, the nondetection
of the other component could have been due to variability.
Similarly, when we found a single radio source in
all eight new epochs, it was not immediately clear whether one component
is always radio-bright and the other radio-faint. 
This ambiguity was resolved by the full astrometric fit, taking
into account parallax, proper motion, as well as orbital motion,
and combining the VLBA and Keck data (Paper~II). The fit is consistent
with a continuous detection of the secondary component and a continuous
nondetection of the primary component. The inclusion of the 2010 observations 
reported by \citet{mcl11} allows us to additionally state that only
the secondary component that was detected with the VLBA.  Whatever
is causing this difference in the radio luminosity of the two components,
it appears to be stable on timescales of years.                       
 
To derive an upper limit for the radio emission of LSPM J1314+1320A,
we have used the full astrometric fit to predict its position at
the various times of the VLBA observations. The nominal positional
error for these predicted positions is between 0.2 and 0.5~mas,
i.e., generally smaller than the synthesized beam size. This allows us to make
a more sensitive image by coadding the different radio epochs, all aligned at the respective position of LSPM J1314+1320A, with a positional error within the half-power synthesized beam width.
Since the position uncertainty for the first epoch in 2010 is slightly larger than
the synthesized beam, we excluded it from this approach.
In the individual epochs, the local 1$\sigma$ noise at the predicted
positions of component A was $21-26$~$\mu$Jy, providing a 3$\sigma$
detection limit for the radio emission of component A of $\lesssim63-78$~$\mu$Jy
per epoch.  With the co-added data from eight epochs, we
obtain an upper limit for LSPM J1314+1320A's radio emission of $\lesssim22$~$\mu$Jy
(3$\sigma$), indicating that LSPM J1314+1320B is brighter than LSPM
J1314+1320A by a factor of $\gtrsim30$ (compared to a factor of $\gtrsim10$
when based on the individual epochs), assuming continuous emission.
 
Even though the stars have masses that are consistent to within $\approx2$\%
(Paper~II), they appear to have vastly differing radio emission. 
One possibility to explain this discrepancy would be
that in spite of otherwise almost identical properties, the rotational
velocities of the two components and/or their prior evolution are significantly different, which could result in different magnetic field configurations. 
Rotation velocity is well known as a main energy reservoir for stellar activity. While differences in rotational velocities (when measured as $v$\,sin$i$)
have been found in several nearby very low mass binaries \citep{kon12},
no such resolved $v$\,sin$i$ measurements have yet been published
for LSPM J1314+1320AB. \citet{mcl11} reported an unresolved measurement
of $v$\,sin$i$=45$\pm$5\,km\,s$^{-1}$, indicative of rapid rotation
in at least one component. Preliminary results from a resolved measurement
indicate that the two stars have very similar $v$\,sin$i$, close
to the unresolved measurement (Williams et al., {\it in prep.}).        
However, it is not clear how large a difference in $v$\,sin$i$ would be required
to explain the factor of $\gtrsim30$ in radio luminosity between
two otherwise almost identical stars. 

Differences in $v$\,sin$i$ could come from different rotational velocities
or different alignments of the rotational axes (or both). 
In the latter case, the differences in emission could also occur if 
the radio emission is highly beamed, as suggested by work on the
electron-cyclotron maser emission which is thought to be occurring
on ultracool dwarfs (e.g., \citealp{hal07,nic12}).  
The question of aligned rotational axes in binaries is an active
area of research (e.g., \citealp{alb14}) with no clear answers.  
Also, radio surveys provide no empirical evidence on how strongly the 
radiation is beamed (e.g., \citealp{mcl12}). 
In either case, very different rotational velocities or different axis alignments
of two otherwise almost identical stars in a coeval binary, would
have interesting implications for the formation and evolution of
very low mass stars.

\section{Upper Limits to the Presence of Planets in Orbit around
LSPM J1314+1320B\label{sec_pla}}                                        
 
\subsection{Assessment of astrometric residuals} 
 
To search for potential signatures of extrasolar planets, we assume
that any residual astrometric trends, after the parallax, proper
motion, the binary orbit have been taken into account,
may be due to the reflex motions from one or more planets. In the absence
of a clear trend, the scatter of the residuals of the astrometric
fit provides significant limits to the presence of planets 
 
In our data, there are no obvious systematic trend in the residuals.
Each step of the Markov chain has an independent
realization of the residuals about that particular astrometric
solution.  For each epoch, we thus calculated the mean and rms of that data
point's residual with respect to all MCMC models.
The residuals in RA and
Declination show no obvious trend as a function of time in
Figure~\ref{fig_1dres}, nor when considering the 2D residuals shown
in Figure~\ref{fig_2dres}, where systematic trends due to reflex
motion could also show up. The 3$\sigma$ rms noise in the 2D residuals
is 0.26~mas, and in RA\,cos\,$\delta$ and Declination it is 0.15~mas and 0.50~mas,
respectively. Overall, these residuals are not unusually high when
compared to our previous observations (e.g., \citealp{fbr13}).          

Depending on the eccentricity of a hypothetical planetary orbit,
its orientation relative to the line of sight, and the spacing of
observations within the orbital period, the observed reflex
motion could lie below these limits. In the following, we primarily
use the 2D residuals to define the limit. In Section~\ref{sub_simu},
we will additionally use simulated planetary orbits and associated
reflex motions to assess the detection completeness.                    
 
\subsection{Discussion of errors} 
 
Realistic VLBA position errors usually exceed the formal position fitting uncertainties. 
To account for this, we introduce ``error floors'' in both RA and Declination 
to represent systematic errors.  These are treated as parameters in the 
full parallax/proper motion/orbit fit (Paper~II), and we find
$\sigma_{\rm \alpha cos \delta}=0.064^{+0.037}_{-0.024}$~mas and 
$\sigma_{\delta}=0.180^{+0.106}_{-0.067}$~mas. 
These error floors should be added in quadrature to those given in Table~\ref{tab_pos}.
 
If the reflex motion due to a planet does not average out over the course 
of the experiment, then the reflex motion signal may contribute to the 
error floors, and as a result the reflex motion will appear less significant than it
actually is. However, the temporal shape of the reflex motion signal
will remains unaffected.  Upper limits on reflex motions are thus doubly
conservative: on the one hand, the residuals contain unmodeled effects
other than planets, and on the other hand the significance of any
reflex motion signal may be underestimated.  
                
In the following analysis, we assume
that the centroid of the radio emission coincides with the centroid
of the stellar photosphere. This would require that the active regions
that are producing the radio emission either fill enough of the stellar
photosphere that the centroid remains stationary even though the
star is rotating or that the active regions are stationary close
to the poles of the star. If there are just a few active regions,
the radio centroid will wobble around the optical/infrared centroid.    
 
Since our full model of the two binary components has produced estimates
of their radii, from a comparison of their masses and their bolometric
luminosities with evolutionary models, we can estimate the maximum amplitude
of this wobble by assuming that the active regions are at the
photospheric radius. The observability of this effect and its timescale
depend on the currently unknown stellar rotation period and the inclination
of the rotation axis. The estimated radius of $r\approx1.8$\,$R_{\rm
jup}$ (Paper II) corresponds to an angular size of 54~$\mu$as at
the distance of LSPM J1314+1320AB.  Thus, asymmetric coronal structures
could thus cause a wobble of $<$54~$\mu$as, which is on the order of the 
RA error floor and thus conservatively accounted for in our analysis.
 
\subsection{Inferred astrometric limits on the presence of planets} 
 
The upper limits to the reflex motion translate into an excluded
region in the phase space of planetary mass and orbital period via
Kepler's Third Law. For a discussion of the limits for circular planetary
orbits, see \citet{eis01a} and \citet{fbr13}. Basically, the reflex motion $\theta$, as a function of orbital radius $a$, is determined by the observationally constrained orbital period $P$ as follows:

\begin{equation}
\theta = \frac{a}{d} \frac{m_c}{m_*} = \frac{m_c}{d} \left( \frac{GP^2}{4\pi^2(m_*+m_c)^2}\right)^{1/3},
\end{equation}

where $d$ is the distance, $m_c$ is the companion mass, and $m_*$ is the mass of the central object. The 2D residuals thus result in an excluded region shown as a shaded area in Figure~\ref{fig_planet}; we note that the
lower RA residuals exclude a subset of orbits with even lower planetary
mass limits, while the limit on reflex motions only in Declination
is higher.  The 3$\sigma$ upper limit from the 2D astrometric residuals
excludes planets of masses 0.7 to 10~$M_{\rm jup}$ for
orbital periods of 600 days to 10 days, respectively.  The limiting planetary mass as a function of orbital period can be described as $M\approx47\times
P(d)^{-2/3}$~$M_{\rm jup}$, where $P(d)$ is the orbital period in
days.  While Figure~\ref{fig_planet} does not show the long periods traced by the inclusion of the 2010 data, no significant deviations from the model are seen on those timescales either (Figures~\ref{fig_1dres} and \ref{fig_2dres}).    
Given our observational scheduling, we are sensitive to nearly continuous range 
of orbital periods from 10 to 1000 days, since for any period in this range
we have at least one pair of measurements that avoid near integer multiples of orbits (shown in Figure~\ref{fig_planet}).
A quantitative assessment of our sensitivity to orbital periods requires simulations of planetary orbits and their reflex motions, which we do 
in Section~\ref{sub_simu}.
 
At small orbital radii, our experiment is probing the outer habitable
zone of LSPM J1314+1320B, which extends to a distance of $\sim0.06$~AU
from the central object (\citealp{sel07}; shown in Figure~\ref{fig_planet}).
Our VLBA astrometry allows us to exclude planets of $\approx7$~$M_{\rm
jup}$ in the outer habitable zone.

\subsection{Assessment of planet detection completeness\label{sub_simu}}
 
To assess the detection completeness to different planetary orbits
in detail, we have simulated orbits over the full range of the 
astrometric model of the binary, making use of the
mass and distance of LSPM J1314+1320B.  We have simulated 10$^6$
planetary orbits at random viewing angles and with eccentricities
of $e=0-0.6$, covering our parameter space in orbital periods and
planetary mass (10--1000 days and 0.1--10~$M_{\rm jup}$). In the
simulated astrometric data, we approximately account for proper motion 
by first subtracting linear fits of the systematic motion 
in both RA and Declination to then assess the residuals,
again separately in RA and Declination. We adopt a simple detection
criterion where the residuals in either RA or Declination have to
be larger than the 3$\sigma$ limits reported above. Finally, we calculated
the 95\% detection fraction from the binned simulation results.         
 
The results of these simulated planetary orbits are overplotted in
Figure~\ref{fig_planet} for circular orbits as well as elliptical
orbits with $e=0.6$. The estimated 95\% detection completeness corresponds
very well with the criterion derived from the 2D 3$\sigma$ astrometric
residuals above, across almost the entire relevant range of orbital
periods, as expected from our observation intervals. Only
at the shortest and longest orbital periods do the limits rise toward
higher masses.  Shorter periods that we
cannot resolve with our observing sequence would show up as extra
astrometric noise throughout the experiment and would be absorbed in
our error floors.  Orbital periods much longer than our spanned observations
would add a constant position offset, and more generally or would add to the
apparent astrometric noise.  Additionally, our setup is slightly less complete 
for elliptical orbits when compared to circular orbits, i.e., their limits generally
lie at higher masses for a given orbital period.                        
 
Note that the simple detection criterion in this test does not account
for the degeneracy of orbital periods of one year with the Earth's
orbit which defines the parallax measurement.
However, it is very unlikely that a planet's reflex motion, measured
at multiple intervals, would match up with the parallax motion in
both RA and Declination. Additional tests with simulated orbits suggest
that this effect only slightly reduces our sensitivity to such planets,
but we refrain from discussing this in more detail since we can rule
out such orbits from orbit stability considerations within this binary
system.                                                                 
 
\subsection{Stability of planetary orbits} 
 
The limit on long orbital periods can be constrained with
information from the full astrometric fit (Paper~II), by including
stability considerations of a planet orbiting one
component of a binary system.  Using the empirical expression for
the critical semimajor axis ($a_c$) of such orbits from \citet{hol99},
above which planetary orbits would not be stable, we obtain $a_c=0.23$~AU,
corresponding to an upper limit of stable planetary orbital periods
of 133 days. While our planetary mass limit based on the absence
of reflex motions is already very low for these long orbital periods
(for example, a planet of 1~$M_{\rm jup}$ and an orbital period of
330 days can be excluded), this constraint rules out a substantial
additional region in the phase space of planetary mass vs orbital
period, as shown in Figure~\ref{fig_planet}. Only the white area
in that figure remains as potential phase space for hypothetical
planets in orbit around LSPM J1314+1320B.                               
 
\subsection{Comparison with other planet-detection methods} 
 
Our observations provide constraints on the presence of extrasolar
planets in a parameter range that is otherwise impossible to access.
The principal comparison here is with radial velocity (RV) surveys
for planets, which are sensitive to $M$sin$i$. The sensitivity of
this method is biased toward close-in planets (e.g., \citealp{eis01b}). While
no RV monitoring of LSPM J1314+1320AB has yet been published, we
now assess what fraction of the parameter space shown in Figure~\ref{fig_planet}
may be accessible with such measurements. The main challenge is the
high $v$\,sin$i$=45\,km\,s$^{-1}$ \citep{mcl12}.                        
 
In an RV survey of ultracool dwarfs, \citet{bla10} reach an RV precision
of 100--300~m\,s$^{-1}$ for slowly rotating stars with $v$sin$i<$30\,km\,s$^{-1}$.
For the 13 ultracool dwarfs in their sample with  $v$sin$i>$30\,km\,s$^{-1}$
(all L dwarfs), the median uncertainty is 390~m\,s$^{-1}$ with a
standard deviation of 285~m\,s$^{-1}$. 
Using this uncertainty to assess the feasibility of RV planet searches
toward LSPM J1314+1320AB, the corresponding 3$\sigma$ limit to the
detection of planets at 1170~m\,s$^{-1}$ is indicated in Figure~\ref{fig_planet}.
A comparison with our constraints from astrometric monitoring reveals
that RV monitoring could produce slightly better constraints
on lower-mass planets at small orbital radii of $\lesssim 0.08$~AU, and,
hence, the two methods are complementary. 
 
Absolute astrometry is not necessary to conduct reflex
motion searches for extrasolar planets.  With relative astrometry
in $I$ band, such a search has been carried out toward 20 ultracool
dwarfs by \citet{sah14}, reaching a median 3$\sigma$ astrometric
residual of 0.60~mas with a standard deviation of 0.16~mas. The 3$\sigma$
2D residual of 0.26~mas that we find toward LSPM J1314+1320B thus
is significantly better, but at the same time a larger sample of
targets is more easily available in the near-infrared than in the
radio range.

\section{Summary and conclusions\label{sec_sum}} 
 
We presented multi-epoch VLBA observations of the ultracool dwarf
binary LSPM J1314+1320AB. In combination with Keck imaging data,
a complete orbit solution was obtained which is discussed in Paper~II.
In this paper, we have used the VLBA data and the orbit solution
to determine the radio properties of both binary components, a parallax,
proper motions, and astrometric limits to the presence of extrasolar
planets. Our conclusions can be summarized as follows:                  
 
\begin{itemize} 
\item The radio emission observed on milliarcsecond scales toward
LSPM J1314+1320AB is entirely due to its secondary component (B),
which is detected at an average flux density of 666~$\mu$Jy.
An upper limit of $\lesssim22$~$\mu$Jy\,bm$^{-1}$ (3$\sigma$) 
is derived for continuous radio emission of the primary (A). Given that
our full astrometric fit yields masses for the two components that
are consistent to within $\approx2$\% (Paper II), this difference
by a factor of $\gtrsim30$ is remarkable. This could point
to differences in stellar rotation as an underlying factor governing
stellar activity, or to differences in the evolution of stellar rotation over time, both potentially resulting in different magnetic field configurations. Alternatively, the radio
emission could be beamed  toward us for the secondary but not for the primary if the rotation axes of the two stars are not aligned. Resolved $v$\,sin$i$ measurements of the two stars could potentially elucidate these issues (Williams et al., {\it in prep.}).                                         
\item The VLBA data allow us to measure the absolute parallax toward
LSPM J1314+1320AB as $\pi_{\rm abs}=57.975\pm0.045$~mas, with
an uncertainty 60 times lower than previous measurements and corresponding
to a distance of $d=17.249\pm0.013$~pc.
Also we obtained an absolute proper motion, which is similar, but more accurate than, 
the previous estimate. Our absolute astrometry will constitute an important
cross-check for the astrometric calibration of {\it Gaia}.                
 
\item Given the full orbit solution and characterization of the binary
components in Paper~II, we can assess the impact of marginally resolved
coronal structures on the centroid
of the radio emission. The estimated radius of the secondary, and
thus the maximum amplitude of any astrometric wobble due
to asymmetric coronal structures, corresponds to an angular size
of 54~$\mu$as at the distance of LSPM J1314+1320AB. This is on the
order of the lowest derived (RA) error floor and thus conservatively accounted for in our analysis.                                                        
 
\item We find no obvious signature of reflex motions due to the
presence of one or more extrasolar planets in orbit around LSPM J1314+1320B.
A 3$\sigma$ limiting planetary mass as a function of orbital period
is given as $M(M_{\rm jup})=47\times P(d)^{-2/3}$ (where $P(d)$ is
the orbital period in days), ruling out, for example, a planet of
2.2~$M_{\rm jup}$ with an orbital period of 100~days. Applying our
full orbit solution for the binary, we additionally obtain a general
upper limit for orbital radii ($a\lesssim0.23$~AU) of stable
planetary orbits in this binary system.  Most of the parameter range of the
derived constraints is inaccessible to RV surveys, but they do
add additional limits on lower-mass planets at small orbital radii
of $\lesssim 0.08$~AU.                                                  
 
\item Finally, the combination of VLBA and Keck data has proven to
be very powerful in this project. The VLBA data provide excellent
absolute astrometry of, in this case, one binary component, and the
Keck data provide excellent relative astrometry of both components.
In combination, the two datasets provide the basis for a full astrometric
fit to an ultracool-dwarf binary at unprecedented accuracy.             
\end{itemize}

\acknowledgments 
The National Radio Astronomy Observatory is a facility of the National
Science Foundation operated under cooperative agreement by Associated
Universities, Inc.                                                      
JF acknowledges helpful discussions with Andreas Brunthaler, Peter
Williams, Alex Wolszczan, Andreas Seifahrt, and Elisabeth Newton.                       
 
 
 
{\it Facilities:} \facility{VLBA}, \facility{Keck}. 
 
 
 
 
 
\bibliographystyle{apj.bst} 
\bibliography{2m1314}

\begin{thebibliography}{}
\expandafter\ifx\csname natexlab\endcsname\relax\def\natexlab#1{#1}\fi

\bibitem[{{Albrecht} {et~al.}(2014){Albrecht}, {Winn}, {Torres}, {Fabrycky},
  {Setiawan}, {Gillon}, {Jehin}, {Triaud}, {Queloz}, {Snellen}, \&
  {Eggleton}}]{alb14}
{Albrecht}, S., {Winn}, J.~N., {Torres}, G., {et~al.} 2014, \apj, 785, 83

\bibitem[{{Baraffe} {et~al.}(2015){Baraffe}, {Homeier}, {Allard}, \&
  {Chabrier}}]{bar15}
{Baraffe}, I., {Homeier}, D., {Allard}, F., \& {Chabrier}, G. 2015, \aap, 577,
  A42

\bibitem[{{Berger}(2002)}]{ber02}
{Berger}, E. 2002, \apj, 572, 503

\bibitem[{{Berger} {et~al.}(2001){Berger}, {Ball}, {Becker}, {Clarke}, {Frail},
  {Fukuda}, {Hoffman}, {Mellon}, {Momjian}, {Murphy}, {Teng}, {Woodruff},
  {Zauderer}, \& {Zavala}}]{ber01}
{Berger}, E., {Ball}, S., {Becker}, K.~M., {et~al.} 2001, \nat, 410, 338

\bibitem[{{Blake} {et~al.}(2010){Blake}, {Charbonneau}, \& {White}}]{bla10}
{Blake}, C.~H., {Charbonneau}, D., \& {White}, R.~J. 2010, \apj, 723, 684

\bibitem[{{Bower} {et~al.}(2009){Bower}, {Bolatto}, {Ford}, \& {Kalas}}]{bow09}
{Bower}, G.~C., {Bolatto}, A., {Ford}, E.~B., \& {Kalas}, P. 2009, \apj, 701,
  1922

\bibitem[{{Dressing} \& {Charbonneau}(2015)}]{dre15}
{Dressing}, C.~D., \& {Charbonneau}, D. 2015, \apj, 807, 45

\bibitem[{{Dupuy} {et~al.}(2016){Dupuy}, {Forbrich}, {Rizzuto}, {XY}, {XY},
  {XY}, \& {XY}}]{dfr16}
{Dupuy}, T.~J., {Forbrich}, J., {Rizzuto}, A., {et~al.} 2016, \apjl, {\it
  subm.}

\bibitem[{{Dupuy} {et~al.}(2015){Dupuy}, {Liu}, {Leggett}, {Ireland}, {Chiu},
  \& {Golimowski}}]{dup15}
{Dupuy}, T.~J., {Liu}, M.~C., {Leggett}, S.~K., {et~al.} 2015, \apj, 805, 56

\bibitem[{{Eisner} \& {Kulkarni}(2001{\natexlab{a}})}]{eis01b}
{Eisner}, J.~A., \& {Kulkarni}, S.~R. 2001{\natexlab{a}}, \apj, 561, 1107

\bibitem[{{Eisner} \& {Kulkarni}(2001{\natexlab{b}})}]{eis01a}
---. 2001{\natexlab{b}}, \apj, 550, 871

\bibitem[{{European Space Agency}(2016)}]{gaia_sp}
{European Space Agency}. 2016, {Gaia: Science Performance},\\
  \url{http://www.cosmos.esa.int/web/gaia/science-performance}

\bibitem[{{Forbrich} \& {Berger}(2009)}]{fbe09}
{Forbrich}, J., \& {Berger}, E. 2009, \apjl, 706, L205

\bibitem[{{Forbrich} {et~al.}(2013){Forbrich}, {Berger}, \& {Reid}}]{fbr13}
{Forbrich}, J., {Berger}, E., \& {Reid}, M.~J. 2013, \apj, 777, 70

\bibitem[{{Hallinan} {et~al.}(2007){Hallinan}, {Bourke}, {Lane}, {Antonova},
  {Zavala}, {Brisken}, {Boyle}, {Vrba}, {Doyle}, \& {Golden}}]{hal07}
{Hallinan}, G., {Bourke}, S., {Lane}, C., {et~al.} 2007, \apjl, 663, L25

\bibitem[{{Holman} \& {Wiegert}(1999)}]{hol99}
{Holman}, M.~J., \& {Wiegert}, P.~A. 1999, \aj, 117, 621

\bibitem[{{Kirkpatrick} {et~al.}(1997){Kirkpatrick}, {Henry}, \&
  {Irwin}}]{kir97}
{Kirkpatrick}, J.~D., {Henry}, T.~J., \& {Irwin}, M.~J. 1997, \aj, 113, 1421

\bibitem[{{Konopacky} {et~al.}(2012){Konopacky}, {Ghez}, {Fabrycky},
  {Macintosh}, {White}, {Barman}, {Rice}, {Hallinan}, \& {Duch{\^e}ne}}]{kon12}
{Konopacky}, Q.~M., {Ghez}, A.~M., {Fabrycky}, D.~C., {et~al.} 2012, \apj, 750,
  79

\bibitem[{{Law} {et~al.}(2006){Law}, {Hodgkin}, \& {Mackay}}]{law06}
{Law}, N.~M., {Hodgkin}, S.~T., \& {Mackay}, C.~D. 2006, \mnras, 368, 1917

\bibitem[{{L{\'e}pine} \& {Shara}(2005)}]{lep05}
{L{\'e}pine}, S., \& {Shara}, M.~M. 2005, \aj, 129, 1483

\bibitem[{{L{\'e}pine} {et~al.}(2009){L{\'e}pine}, {Thorstensen}, {Shara}, \&
  {Rich}}]{lep09}
{L{\'e}pine}, S., {Thorstensen}, J.~R., {Shara}, M.~M., \& {Rich}, R.~M. 2009,
  \aj, 137, 4109

\bibitem[{{McLean} {et~al.}(2011){McLean}, {Berger}, {Irwin}, {Forbrich}, \&
  {Reiners}}]{mcl11}
{McLean}, M., {Berger}, E., {Irwin}, J., {Forbrich}, J., \& {Reiners}, A. 2011,
  \apj, 741, 27

\bibitem[{{McLean} {et~al.}(2012){McLean}, {Berger}, \& {Reiners}}]{mcl12}
{McLean}, M., {Berger}, E., \& {Reiners}, A. 2012, \apj, 746, 23

\bibitem[{{Melis} {et~al.}(2014){Melis}, {Reid}, {Mioduszewski}, {Stauffer}, \&
  {Bower}}]{mel14}
{Melis}, C., {Reid}, M.~J., {Mioduszewski}, A.~J., {Stauffer}, J.~R., \&
  {Bower}, G.~C. 2014, Science, 345, 1029

\bibitem[{{Nichols} {et~al.}(2012){Nichols}, {Burleigh}, {Casewell}, {Cowley},
  {Wynn}, {Clarke}, \& {West}}]{nic12}
{Nichols}, J.~D., {Burleigh}, M.~R., {Casewell}, S.~L., {et~al.} 2012, \apj,
  760, 59

\bibitem[{{Reid} \& {Brunthaler}(2004)}]{reb04}
{Reid}, M.~J., \& {Brunthaler}, A. 2004, \apj, 616, 872

\bibitem[{{Reid} \& {Honma}(2014)}]{reh14}
{Reid}, M.~J., \& {Honma}, M. 2014, \araa, 52, 339

\bibitem[{{Sahlmann} {et~al.}(2016){Sahlmann}, {Lazorenko}, {Bouy},
  {Mart{\'{\i}}n}, {Queloz}, {S{\'e}gransan}, \& {Zapatero Osorio}}]{sah16}
{Sahlmann}, J., {Lazorenko}, P.~F., {Bouy}, H., {et~al.} 2016, \mnras, 455, 357

\bibitem[{{Sahlmann} {et~al.}(2014){Sahlmann}, {Lazorenko}, {S{\'e}gransan},
  {Mart{\'{\i}}n}, {Mayor}, {Queloz}, \& {Udry}}]{sah14}
{Sahlmann}, J., {Lazorenko}, P.~F., {S{\'e}gransan}, D., {et~al.} 2014, \aap,
  565, A20

\bibitem[{{Schlieder} {et~al.}(2014){Schlieder}, {Bonnefoy}, {Herbst},
  {L{\'e}pine}, {Berger}, {Henning}, {Skemer}, {Chauvin}, {Rice}, {Biller},
  {Girard}, {Lagrange}, {Hinz}, {Defr{\`e}re}, {Bergfors}, {Brandner},
  {Lacour}, {Skrutskie}, \& {Leisenring}}]{sch14}
{Schlieder}, J.~E., {Bonnefoy}, M., {Herbst}, T.~M., {et~al.} 2014, \apj, 783,
  27

\bibitem[{{Selsis} {et~al.}(2007){Selsis}, {Kasting}, {Levrard}, {Paillet},
  {Ribas}, \& {Delfosse}}]{sel07}
{Selsis}, F., {Kasting}, J.~F., {Levrard}, B., {et~al.} 2007, \aap, 476, 1373

\bibitem[{{Williams} {et~al.}(2015){Williams}, {Berger}, {Irwin},
  {Berta-Thompson}, \& {Charbonneau}}]{wil15}
{Williams}, P.~K.~G., {Berger}, E., {Irwin}, J., {Berta-Thompson}, Z.~K., \&
  {Charbonneau}, D. 2015, \apj, 799, 192

\bibitem[{{Wolszczan} \& {Route}(2014)}]{wol14}
{Wolszczan}, A., \& {Route}, M. 2014, \apj, 788, 23

\end{thebibliography}

\end{document}